\documentclass[twocolumn]{cinc}
\usepackage{graphicx}
\usepackage{amsmath}
\usepackage{amsfonts}

\begin{document}
\bibliographystyle{cinc}

\title{Multilabel 12-Lead Electrocardiogram Classification Using \\
Gradient Boosting Tree Ensemble}

\author {Alexander W Wong$^{1}$, Weijie Sun$^{2}$, Sunil V Kalmady$^{2}$, Padma Kaul$^{2}$, Abram Hindle$^{1}$\\
\ \\
 $^1$ University of Alberta, Edmonton, Canada \\
$^2$ Canadian VIGOUR Centre, Edmonton, Canada }

\maketitle

\newcommand{\officialvalscore}{{0.476} }
\newcommand{\officialtestscore}{{-0.080} }

\begin{abstract}

The 12-lead electrocardiogram (ECG) is a commonly used tool for detecting cardiac abnormalities such as atrial fibrillation, blocks, and irregular complexes.
For the PhysioNet/CinC 2020 Challenge, we built an algorithm using gradient boosted tree ensembles fitted on morphology and signal processing features to classify ECG diagnosis.

For each lead, we derive features from heart rate variability, PQRST template shape, and the full signal waveform.
We join the features of all 12 leads to fit an ensemble of gradient boosting decision trees to predict probabilities of ECG instances belonging to each class.
We train a phase one set of feature importance determining models to isolate the top 1,000 most important features to use in our phase two diagnosis prediction models.
We use repeated random sub-sampling by splitting our dataset of 43,101 records into 100 independent runs of 85:15 training/validation splits for our internal evaluation results.

Our methodology generates us an official phase validation set score of \officialvalscore and test set score of \officialtestscore under the team name, CVC, placing us 36 out of 41 in the rankings.

\end{abstract}

\section{Introduction}

The electrocardiogram (ECG), when correctly interpreted, is an effective tool for detecting cardiac diseases.
Despite much research in computerized interpretations of ECGs, trained human over-reading and confirmation is required and emphasized in published reports~\cite{SMITH201988,MADIAS2018413}.
This work classifies standard 12-lead ECGs to their clinical diagnosis as part of the \emph{PhysioNet/CinC 2020 Challenge}~\cite{physionet_challenge_2020}.
We develop a multi-label classification algorithm using entropy and signal processing inspired features and a gradient boosting decision tree ensemble.

\subsection{Dataset \& Scoring Criteria}

The official phase dataset contains a total of 43,101 ECG records.
Each record contains a set of one or more SNOMED CT codes, with only a subset of 27 codes evaluated in the challenge.
The challenge objective is to maximize the metric: $\sum_{ij} w_{ij} a_{ij}$.
Given a set of diagnoses $C = \{c_i\}$, we compute a confusion matrix $A = [a_{ij}]$ where $a_{ij}$ contains records that are classified as class $c_i$ and belong to class $c_j$.
The weights $W = [w_{ij}]$, are set by the challenge to indicate clinical similarity between classes.
Refer to Perez Alday \emph{et al.}~\cite{physionet_challenge_2020} for the description of the challenge scoring function weights and ECG dataset.

\section{Methodology}

\begin{figure}[tbp]
  \centering
  \includegraphics[width=7.9cm]{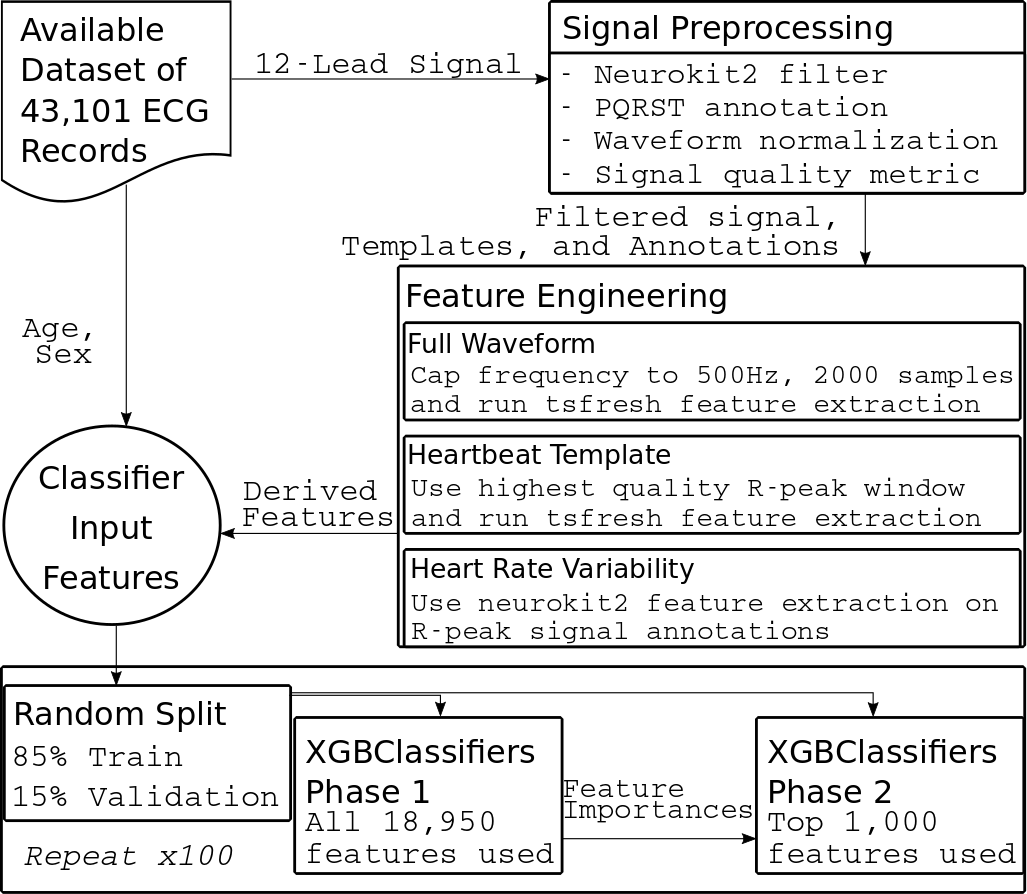}
  \caption{Methodology overview. Feature engineering is performed concurrently for each lead then concatenated.}
  \label{fig:methodology}
\end{figure}

Our approach is inspired by existing methods which use feature engineering and shallow learning classifiers~\cite{goodwin_classification_2017}.
Figure~\ref{fig:methodology} shows an overview of our learning algorithm pipeline from first cleaning and preprocessing the ECG, to then extracting the full waveform, heartbeat template, and heart rate variability features, finally using these features as input to our binary classifiers.

We rely on the \emph{NeuroKit2} (version \texttt{0.0.40}) neurophysiological signal processing library for ECG signal cleaning, PQRST annotation, signal quality calculation, and heart rate variability metrics~\cite{neurokit2}.
We also use the time series feature extraction library \emph{tsfresh} (version \texttt{0.16.0}) for analysis of the PQRST template and the full waveform~\cite{CHRIST201872}.

\subsection{Signal Pre-processing}

First we perform signal pre-processing to normalize and clean the raw ECG signal.
Slow drift and DC offset are removed with a Butterworth highpass filter followed by smoothing using a moving average kernel of 0.02 seconds.
Each of the cleaned leads are independently annotated with the PQRST peaks, the PRT onsets, and PRT offsets.

We isolate one candidate heart beat signal for each lead by segmenting heart beat windows as a $-0.35$ to 0.5 second window around each R-peak, shortening to a $-0.25$ to 0.4 second window if the mean heart rate exceeds 80 beats per minute.
We create an ECG signal quality metric by interpolating the distance of each QRS segment from the average QRS segment in the data.
ECG signal quality is therefore relative for each step in the entire length of the signal, where 1 corresponds to beats that are closest to the average QRS and 0 corresponds to beats that are most distant to the average QRS.
We use the PQRST beat window with the highest signal quality as our candidate lead heartbeat template.

\subsection{Feature Engineering}

Our engineered features are categorized as one of three categories.
Full waveform features are derived using the end-to-end ECG signal.
Template features are constructed from the extracted PQRST window during pre-processing.
Heart rate variability features rely on the relative distances between each R-peak.
Each extraction technique is performed independently per lead and concatenated together prior to classification.

For full waveform and heartbeat template features, we use the cleaned ECG signal and apply the \emph{tsfresh} feature extraction library.
For full waveform features, we cap the signal sampling rate to a maximum of 500Hz before limiting the signal to the middle 2,000 samples to remove starting and trailing artifacts.
Template features are derived from the isolated heart beat window with highest signal quality.
Using the default feature extraction settings, we generate 763 template and 763 full waveform features per lead.
The extracted features include autoregressive model coefficients, change quantiles, aggregate linear least-squares regression trends, peak counts, sample/approximation entropy, energy, continuous waveform transform coefficients, fast fourier transform coefficients, and other descriptive statistics of the signal.

Heart rate variability (HRV) features are generated from the cleaned signal and corresponding R-peak annotations using \emph{NeuroKit2}.
We use the default feature extraction settings and generate 53 different HRV features per lead.
HRV features include: mean, median, standard/absolute deviation, and interquartile range of the RR intervals; standard deviation of the successive differences between RR intervals; proportion of RR intervals greater than 50/20ms over total RR intervals; and geometric indices measuring triangular interpolation of the RR interval distribution.

For each 12-lead record we combine all three categories of engineered features with the age and sex parsed from the ECG record metadata.
We arrive at a $12 \cdot (763 + 763 + 53) + 2 = 18,950$ length feature vector per 12-lead record.

\subsection{Classification}

We train a XGBoost binary classifier for each of the 27 clinical diagnoses, using \texttt{xgboost@1.1.1}~\cite{chen_xgboost_2016}.
We sample each training instance with a selection probability proportional to the regularized absolute value of the gradients.
Early stopping is set to 20 rounds with binary logistic regression as our objective function.

We use the evaluation scoring weights as instance sample weights, capping positive examples to a 0.5 threshold.
For example, when training the 1st degree atrioventricular block (IAVB) classifier we consider instances of bradycardia (Brady), incomplete right bundle branch block (IRBBB), prolonged PR interval (LPR), sinus arrhythmia (SA), and sinus bradycardia (SB) as positive examples with 0.5 weight.
Other labels that have scoring function weights below 0.5 are treated as negative examples with a sample weight of 1.
To account for the dataset label imbalance, we further scale the positive weight using the number of negative samples over the positive samples in the training set split.

Our classification models are trained in two phases.
First, we randomly sub-sample our total dataset, splitting our 43,101 records into an 85:15 training/validation set split.
In the phase one, we train using all 18,950 features to estimate the feature importances.
Feature importance is defined as the model reported gain in accuracy contributed by the feature over all branches in the decision tree.
We average the importances outputted by the 27 binary classifiers to get the mean importance for each feature.
We rank all of the features by their mean importance and keep the top 1,000 important features.
In phase two, we train new models using the same training and validation split but limiting the classifier input to the top 1,000 most important features.
This process is repeated 100 times, exhausting our available dataset.

For the submission component of the competition, we omit training phase one of our classification models due to insufficient computing resources and time constraints.
We overcome this limitation by using the phase one models that we trained locally.
All 100 phase one model feature importances are averaged together to generate a overall mean feature importance, using our entire available dataset.
Our challenge submission's classification model only needs to train the phase two set of classifiers, using the top 1,000 features that we computed as a prior.

\section{Results}

\begin{figure}[ht]
  \centering
  \includegraphics[width=7.9cm]{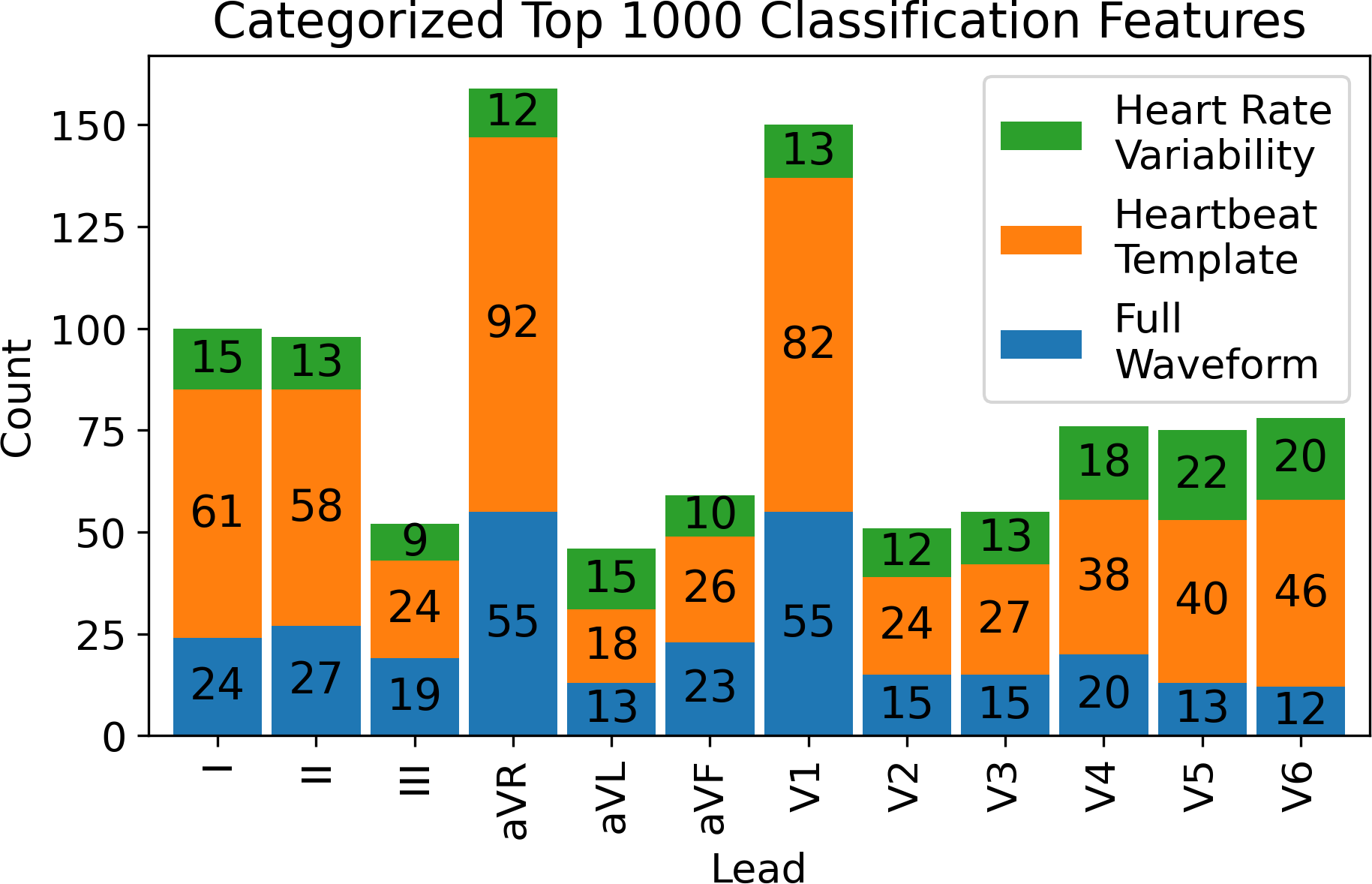}
  \caption{Count of lead and feature categories comprising the top 1,000 features. Age, but not sex, is important meta.}
  \label{fig:top_features}
\end{figure}

A categorical visualization of the top 1,000 features used in the challenge submission model, grouped by lead and feature type, is shown in Figure~\ref{fig:top_features}.
Most of the features are derived from leads \texttt{aVR} (159 features) and \texttt{V1} (150 features).
The \texttt{aVL} lead is least represented with only 46 derived features used by the phase 2 classifier.
The heartbeat template category containing 536 features is the most numerous feature type.
There are 291 full waveform features and 172 heart rate variability features.
The age meta feature parsed from the ECG record header is also used.

\begin{figure}[hb]
  \centering
  \includegraphics[width=7.9cm]{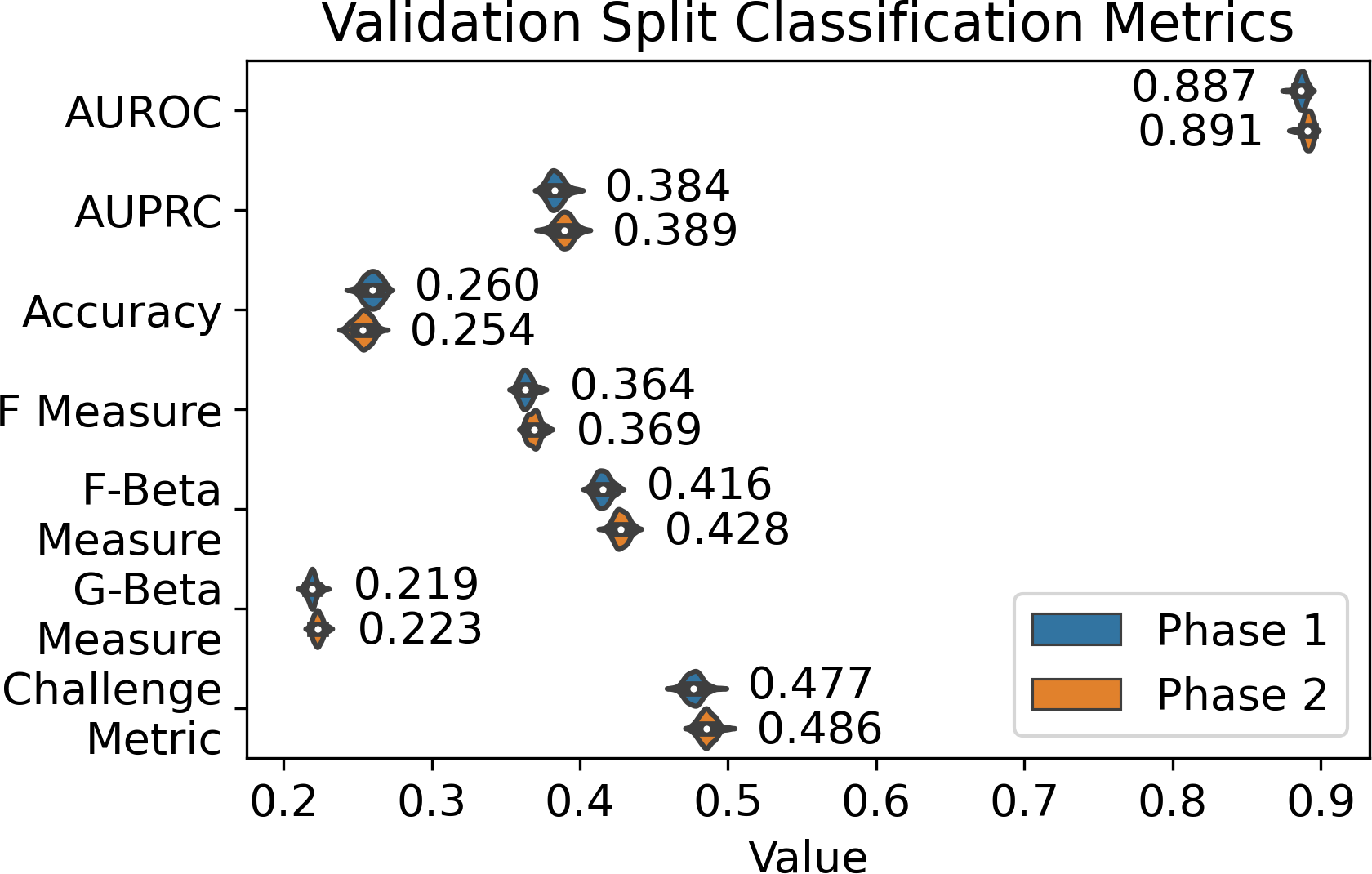}
  \caption{Summary of classification metrics over 100 experiments on all labels. Annotations indicate mean value.}
  \label{fig:classification_metrics_summary}
\end{figure}

We present our metrics from the 15\% validation splits of the dataset for phases one and two of our classification models in Figure~\ref{fig:classification_metrics_summary}.
Our phase two models have higher mean values for all classification metrics except for accuracy.
Using the smaller set of features, our classification metric variances are more closely centered around the mean.
Our methodology attains a phase two mean challenge metric score of 0.486.
Additionally, we attain phase two mean values for AUROC of 0.891, AUPRC of 0.389, accuracy of 0.254, overall $\text{F}_1$ score of 0.369, $\text{F}_\beta$ of 0.428, and $\text{G}_\beta$ measure of 0.223 using $\beta = 2$.

\begin{figure*}[ht]
  \centering
  \includegraphics[width=17.0cm]{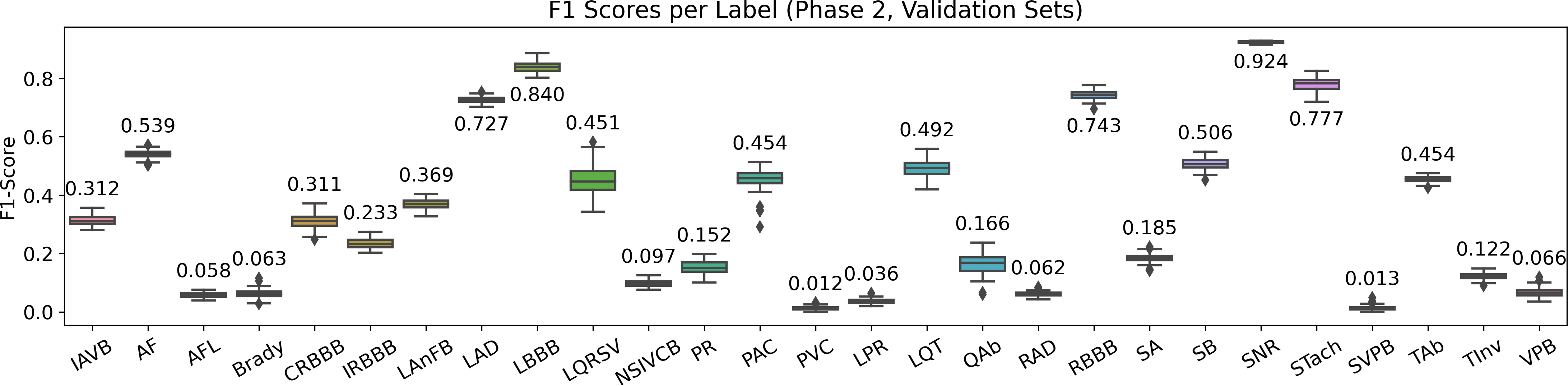}
  \caption{Phase two label-wise validation set $\text{F}_1$ scores over 100 independent runs. Annotations indicate mean value.}
  \label{fig:f1_score}
\end{figure*}

Our model's top three best classified labels are normal sinus rhythm (SNR, $\bar{\text{F}}_1$: 0.924), left bundle branch block (LBBB, $\bar{\text{F}}_1$: 0.840), and sinus tachycardia (STach, $\bar{\text{F}}_1$: 0.777).
A summary of our phase two $\text{F}_1$ scores on the 100 validation splits for each label is shown in Figure~\ref{fig:f1_score}.

Furthermore, we run a Pearson correlation coefficient test between the label $\text{F}_1$ means and the label counts within our dataset.
The statistical test reveals a Pearson correlation coefficient of 0.602 at a p-value of $9.0 \cdot 10^{-4}$.
This result suggests that a positive linear correlation exists between the label occurrence in our dataset and our classification model's $\text{F}_1$ score.

Our methodology achieves a challenge score of \officialvalscore on the official validation set and \officialtestscore for the official test set, ranking team CVC at 36 of 41 teams.

\section{Discussion \& Future Work}

Despite the label specific scaling of our dataset, the correlation between the label occurrence with the $\text{F}_1$ scores suggest further improvements are necessary to mitigate label imbalance.
The label imbalance may be addressed by adding more low occurrence disorders into the existing corpus of ECG records.
Synthesizing new records of low occurrence disorders to use as training data may also prove promising.
Additionally, exploration of new features to use as classifier inputs may reveal common characteristics of specific heart disorders that are currently missing.

Our approach, although applicable to 12-lead ECGs, perform feature extraction on each lead separately before concatenating the features together for classification.
We believe that further improvements can be made utilizing feature extraction approaches capable of handling multi-dimensional time series data.

Our approach does not use additional external datasets, nor do we modify any of the labels provided in the available dataset.
We anticipate that further corrections in the ECG diagnosis labels, and including more ECG records, would enable our approach to achieve more competitive competition scores.

We acknowledge that our internal results and corresponding figures report optimistic values for the classification metrics, as our internal split of the dataset does not include a hold-out test set.
We rely on the hold out test set from the challenge organizers to fairly evaluate our challenge score.
Future work includes replicating our method using a local training, validation, and test set split, reporting label-wise $\text{F}_1$ on the test set.

The requirement of the challenge to train a model on a hold out training set added additional engineering complexity that could not be fully addressed in our final submission.
The computation time of training the phase one feature importance models exceeded the allocated time constraints set by the challenge, using their provided cloud virtual machines.
Our workaround therefore relies on the feature importances generated locally, using the available, released data.
The feature importances used for the challenge submission model may not match the distribution of feature importances of the hold out training set.

\section{Conclusion}

We create an algorithm for the classification of 27 heart conditions using signal processing inspired feature engineering and an XGBoost tree ensemble classifier.
We combine a set of 18,950 features from full waveform, heartbeat template, and heart rate variability groups.
Using 100 repeated random sub-sampling of 85:15 train/validation, we train models to get feature importances and distilled out 1,000 most important features.
Using this reduced set of 1,000 features, we retrain our models and achieve a mean challenge score of 0.486 on our validation split.
For our team, \emph{CVC}, the official phase challenge scores are \officialvalscore on the validation set and \officialtestscore on the test set.
We attain a rank of 36 of 41 qualifying teams.

\section*{Acknowledgments}
We would like to thank Eric Ly and Leiah Luoma of the Canadian VIGOUR Center for their help and guidance during our research journey.

\bibliography{main}

\begin{correspondence}
Alexander W. Wong\\
Department of Computing Science\\
2-32 Athabasca Hall, University of Alberta\\
Edmonton, Alberta, Canada\\
T6G 2E8\\
alex.wong@ualberta.ca
\end{correspondence}

\end{document}